\documentclass[12pt,a4paper]{article}
\usepackage[T2A]{fontenc}
\usepackage[cp1251]{inputenc}
\def\v{\vskip 0.3 cm}
\begin{document}
\def\d{\displaystyle}
\centerline {\bf Globular Clusters as Candidates for Gravitational Lenses }

\centerline {\bf to Explain Quasar--–Galaxy Associations}
\v
\centerline {\bf Yu. L. Bukhmastova}

\vskip0.5cm
\centerline {\it Astronomical Institute, St. Petersburg State University, Russia,}
\v
\centerline {\it bukh@astro.spbu.ru}

\v 
{\bf Keywords:}\hskip 0.2cm {\it quasar---galaxy associations, gravitational lensing, globular clusters, hidden halomass}.
\v
\v
{ INTRODUCTION}
\v
What the nature of the association between some quasars and nearby galaxies is has remained an open question for more than three decades. The essence of this problem is that some quasars are observed in the immediate vicinity of galaxies, with the redshifts of the galaxies being much lower than those of the quasars close to them. A number of publications are devoted to this problem. Among these, we note the monographs on gravitational lensing by Bliokh and Minakov (1989) and Schneider et al. (1992) and the review article by Zakharov (1997).

Arp was among the first to gave rapt attention to close quasar--galaxy pairs. In his paper (Arp 1987), he provided examples of four galaxies at $z\sim 0.01$ that have two or three quasars at a distance of $2^{\prime}$ from the galactic center. These quasars have $z\sim 1$ and are brighter than $m_v\sim 20.$ Since the density of quasars with $m_v$  brighter than 20 is 20 objects per square degree, it seems unlikely to detect two or three quasars around any point. According to the estimates by Schneider et al. (1992), the probabilities of detecting two and three such quasars at an angular distance of $2^{\prime}$ are 0.07 and 0.005, respectively. Two major strategies of searching for quasar---galaxy  pairs are being discussed in the literature (Schneider et al. (1992)).

 The first strategy of searching for such pairs is to consider quasars around galaxies and to investigate the possible excess of the density of quasars around galaxies above the mean quasar density.

The overlap between the Second Reference Catalogue of Bright Galaxies containing 4364 objects and the catalog of quasars (Hewitt and Burbidge 1980) containing 1356 objects showed a statistical significance of the correlation between the sky projections of quasars at different redshifts and bright galaxies at $z<0.05$ (Chu et al. 1984). However, the result was questioned, because the catalog included mostly radio--active quasars. Moreover, analysis of the catalogs of quasars and galaxies (Nieto and Seldner 1982) revealed no statistical significance for associations. Nieto (1977, 1978, 1979) and Nieto and Seldner (1982) found no significant excess of the density of all types of quasars around nearby galaxies but revealed an excess of the density among optically variable quasars.

The second strategy of searching for pairs is to consider galaxies around quasars to determine whether there is an excess of the density of galaxies around quasars above the mean galaxy density. Rozyczka (1972) pointed out that there is such an excess of galaxies around nearby quasars and that low-redshift quasars are located in groups of galaxies at the same redshift. Tyson (1986) found an excess of the density of galaxies around low- and high-redshift
quasars.

Burbidge et al. (1990) conducted a computeraided search for close quasar---galaxy pairs with an angular separation of less than $10^{\prime}.$ As a result of thissearch, a catalog of 577 quasars and 500 galaxies was published. The observed number of associations is several times larger than the expected number of chance coincidences. For example, the probability of detecting a randomly thrown object at an angular distance of $2^{\prime}$ from a galaxy brighter than $15^m$ is estimated to be $10^{-3}.$ Thus, the expected number of chance pairs for 5000 observed quasars is 5, while the catalog contains 38 such pairs. Such a simple estimate indicates that the physical association between quasars and galaxies in pairs is real. This association is also supported by the angular separation--galaxy redshift diagram constructed for 392 pairs. According to this observational relation, the mean linear galaxy---quasar separation in projection onto the sky, $\sim 100$~kpc, remains constant. These data led Burbidge et al. (1990) to conclude that quasars tend to be located in the halos of normal galaxies much more often than is expected for chance projections and that this physical association requires an explanation.

Based on the LEDA catalog of galaxies (Paturel 1997) containing some 100 000 objects and on the catalog of quasars and active galactic nuclei (AGNs) (Veron-Cetty and Veron 1998), we previously (Bukhmastova 2001) searched for close quasar---galaxy pairs. The search for pairs was conducted in such a way that the linear separation between the galaxy and the quasar projection onto the galactic plane did not exceed 150~kpc. The following
criteria were used:

(1) The spatial coordinates $\alpha,\delta$ and $z$ are available
for quasars and galaxies and the magnitudes $m_Q$ are
known for quasars;

(2) The quasar must be farther than the galaxy
(i.e., $z_Q > z_G$);

(3) The quasars must be projected onto the halos
of galaxies up to 150~kpc in size;

(4) The galaxy must have $z_G > 4\cdot 10^{-4}.$

77 483 galaxies and 11 358 quasars satised criterion (1). 1054 galaxies and 3164 quasars were in pairs, which amounted to 8382 quasar---galaxy pairs. Below, such pairs selected by using criteria (2)--(4) are called quasar---galaxy associations.

A statistical analysis of such associations based on computer simulations and on theoretical calculations indicates that their observed number is much larger than the number of chance projections (1200 pairs), suggesting a physical association between the galaxy
and quasar images.

When analyzing the catalog of close quasar---galaxy pairs (Burbidge et al. 1990), Baryshev and Ezova (1997) found the number of quasar---galaxy associations to depend on $a=z_G/z_Q,$ where $z_G$ and $z_Q$ are the redshifts of the galaxy and the quasar from each pair. According to this dependence, galaxies in pairs tend to be located either close to the observer (so that $a<0.1$) or close to the quasar ($a>0.9$) and avoid the central position in the observer---quasar segment.

Based on new data for 8382 pairs, we previously (Bukhmastova 2001) analyzed the dependence of the number of pairs on $a=z_G/z_Q.$ This dependence confermed the previous result that there is an excess of pairs with $a < 0.1$ and $a > 0.9.$ There are at least two hypotheses that can explain the appearance of close quasar--galaxy pairs. One of these is that quasars in associations are in the halos of nearby galaxies; thus, their redshifts are not distance indicators for these objects. Arp (1966, 1987, 1990) adhered to this hypothesis. The second hypothesis is that quasars in associations are only projected onto the halos of nearby galaxies; their redshifts remain cosmological in origin and the enhanced visible luminosity of such quasars is the result of gravitational lensing by objects in the galactic halos. Here, we hypothesize that at least some of such pairs result from gravitational lensing. If this is the case, then the quasars in associations are the images of distant sources. The following question naturally arises: what astrophysical object for each pair acts as the source and what object is the gravitational lens? Canizares (1981) was the first to attempt to explain the appearance of quasar---galaxy associations as due to gravitational lensing by galactic halo stars (microlensing). Thus, the individual halo stars falling on the line of sight acted as the lenses. Faint quasars were chosen as the sources; because of the low surface density of faint quasars, it was subsequently shown that the hypothesis of gravitational lensing could not explain the observed effect. It is important to note that microlensing makes it possible to weakly amplify the source images.

If the gravitational lenses are not stars but globular clusters (GCs), dwarf galaxies, or clustered hidden mass objects with masses $10^5-10^9 M_\odot$ (''mesolenses''), then it becomes possible to strongly amplify the sources (Baryshen and Ezova 1997). The King model is applicable to such objects. This model well fits the density profile in GCs, dwarf galaxies, and, possibly, other self-gravitating hidden mass objects. These objects are assumed to be located in the halos of galaxies in associations.

The existence and abundance of such hidden mass objects in galactic halos were discussed by Paolis et al. (1998). Dalal and Kochanek (2001) noted that the hidden mass in galactic halos could be clustered. The authors investigated seven radio sources with four-component images, in which they detected gravitational lensing. Six lensed sources exhibited an anomalous flux that could be explained by the presence of a  substructure. Analysis showed that the masses of the objects of the gravitational lens substructure lay within the range characteristic of GCs, $10^6-10^9 M_\odot.$

The nature of the gravitational lenses responsible for the appearance of quasar--galaxy associations and for the assumed enhancement of the quasar flux can differ widely. The gravitational lenses can be made up of dark baryonic matter. In his review article, Carr (1994) discussed in detail the various components of the dark baryonic matter. Snowballs produced through the condensation of cold hydrogen, brown dwarfs, Jupiter--like objects composed of hydrogen, dim red stars (Њ dwarfs), clouds of molecular hydrogen, neutron stars, black holes produced by the collapse of baryonic matter, and supermassive black holes can be these components.

The possibility that the gravitational lenses could be formed out of nonbaryonic cold dark matter (CDM) is not ruled out. The need for this mass component in the Universe follows from the prediction of critical total matter density $\Omega_{tot}=1$ in the influationary model and from the prediction of baryonic matter density $\Omega_{bar} = 0.05h^{-2}_{65}$ in the standard model of primordial nucleosynthesis. Thus, $95\%$ of the mass in the Universe can be in the mysterious nonbaryonic form. The Einstein cosmological constant or dark energy is currently believed to account for $70\%$ of the dark matter (Peebles and Ratra 2002). The remaining $25\%$ is accounted for by nonbaryonic CDM, for which the thermal energy and pressure can be disregarded compared to the energy of the rest mass. Dark energy represents a uniform background, while CDM is clustered to form structures of all scales (for a detailed review of the present views of dark energy, see Sahni and Starobinsky (2000)).

In the standard cosmological model, CDM is a dominant dynamical factor that governs the formation of galactic halos and the large-scale structure. High-resolution numerical calculations show that the hierarchy of CDM structures extends from the scales of galaxy superclusters down to the halos of individual galaxies composed of fundamental CDM clusters with masses in the range     $10^6-10^8M_\odot$ (Klypin et al.1999; Morr et al. 1999, 2001). Thus, if the clusters with such masses are assumed to account for $0.2\%,$ then, as follows from an estimate of the lensing probability, on the order of several tens of thousands of quasars could be amplified by gravitational lensing (Baryshev and Ezova 1997).

To explain the large number of quasar--galaxy associations, we accept the Barnothy--Tyson hypothesis (Barnothy 1965) that quasars, at least partially, are the gravitationally amplified images of AGNs. Thus, in our model, compact massive objects located in the central regions of active galaxies, such as Seyfert galaxies and radio galaxies of all types, rather than faint quasars act as the sources. The number of these objects is large enough to provide the observed properties of the associations for ampli.cation factors of 3-5 magnitudes. Kormendy (2000a, 2000b), Kormendy and Ho (2000), Ho and Kormendy (2000), and Bartusiak (1998) gave evidence that one might expect the presence of compact massive objects in the central regions of almost all galaxies. This large number of potential sources, naturally, exceeds the probability of occurrence of the lensing effect. Stuart et al. (2002) assumed black holes with masses $>3\cdot10^9 M_{\odot}$ to be located in the central parts of quasars. The authors showed that a third of the distant quasars at $z\sim 6$ could be amplified by gravitational lenses on the line of sight, with the observed flux being enhanced by several tens of times or more. This study is a logical continuation of the studies of the properties for quasar--galaxy associations initiated by Baryshev and Bukhmastova (1997, 2001). We argue that GCs are suitable candidates for gravitational lenses when we deal with quasar--galaxy associations. We discuss GC parameters, in particular, the distribution of GCs in central surface density. We give the possible locations of GC lenses relative to  the centers of the galaxies in associations and analyze the types of galaxies in pairs. We summarize our main conclusions and discuss possible observational tests.
\v
\v

DATA ON GLOBULAR CLUSTERS
\v
The idea to consider GCs as possible candidates for gravitational lenses was first put forward by Barnothy (1974) using the associations of the quasar 3C 455 with the galaxy NGC 7413 and an M15-type GC as an example. GCs as mesolens candidates were also considered by Yushchenko et al. (1998), Ugolnikov (2001), and Kurt and Ugolnikov (2000). Here, to obtain specific quantitative results, we use the parameters of 135 GCs in the Milky Way; thus, the previously used sample of 56 GCs analyzed by Baryshev and Ezova (1997) was expanded. For each of the GCs, we determined the core radius in pc and the central surface density by using the King model. These parameters are used for rough estimates. The list of Milky Way GCs in Table 1 contains the following parameters:
\newpage
\begin{tabular}{|c|c|c|c|c|c|c|}
\hline
ID\hfil &$ c $\hfil&$r_c^{'}$\hfil &$R$\hfil &${\lg M/M_{\odot}}$\hfil &$R_c$\hfil&$\sigma_0$\\
\hline
\hline
NGC 4147 &  1.80 & 0.10 & 18.8 &  4.39 &  0.56 &  0.9\\
NGC 6325 &  2.50 & 0.03 &  9.4 &  4.44 &  0.08 & 30.1\\
NGC 6402 &  1.60 & 0.83 &  8.7 &  5.89 &  2.16 &  2.5\\
TER 9    &  2.50 & 0.03 &  7.8 &  3.32 &  0.07 &  3.3\\
NGC 6864 &  1.88 & 0.10 & 18.4 &  5.43 &  0.55 & 10.4\\
PAL 12   &  1.94 & 0.20 & 18.7 &  3.82 &  1.12 &  5e-2\\
AM 1     &  1.23 & 0.12 & 119.1&  3.69 &  4.28 &  7e-3\\
NGC 3202 &  1.31 & 1.45 &  5.1 &  5.05 &  2.21 &  0.5\\
NGC 5897 &  0.79 & 1.96 & 12.7 &  4.83 &  7.46 &  0.1\\
NGC 6139 &  1.80 & 0.14 & 10.5 &  5.28 &  0.44 &  12.3\\
NGC 6205 &  1.49 & 0.88 &  7.0 &  5.59 &  1.85 &  1.9\\
NGC 6304 &  1.80 & 0.21 &  6.0 &  4.92 &  0.38 &  7.3\\
NGC 6535 &  1.30 & 0.42 &  6.8 &  3.81 &  0.86 &  0.2\\
NGC 6656 &  1.31 & 1.42 &  3.2 &  5.53 &  1.36 &  4.0\\
PAL 13   &  0.66 & 0.48 & 26.3 &  3.14 &  3.79 &  0.2\\
NGC 5986 &  1.22 & 0.63 & 10.3 &  5.48 &  1.95 &  2.0\\
NGC 6723 &  1.05 & 0.94 &  8.6 &  5.15 &  2.43 &  0.9\\
NGC 6316 &  1.55 & 0.17 & 11.5 &  5.49 &  0.59 & 14.5\\
NGC 6712 &  0.90 & 0.94 &  6.7 &  4.98 &  1.89 &  1.6\\
NGC 5139 &  1.24 & 2.58 &  5.1 &  6.38 &  3.95 &  3.8\\
NGC 5634 &  1.60 & 0.21 & 25.3 &  5.18 &  1.59 &  0.9\\
PAL 14   &  0.72 & 1.1  & 72.2 &  3.83 &  23.83&  3e-3\\
NGC 6426 &  1.70 & 0.26 & 19.9 &  4.50 &  1.55 &  0.2\\
NGC 6558 &  2.50 & 0.03 &  6.4 &  4.50 &  0.06 & 74.6\\
NGC 6569 &  1.27 & 0.37 &  8.5 &  5.19 &  0.94 &  4.1\\
NGC 6624 &  2.50 & 0.06 &  7.9 &  5.04 &  0.14 & 42.5\\
NGC 1851 &  2.24 & 0.08 & 12.2 &  5.42 &  0.29 & 27.9\\
AM 4     &  0.50 & 0.42 & 29.2 &  2.41 &  3.68 &  2e-3\\
PAL 15   &  0.60 & 1.21 & 43.6 &  4.10 &  15.83&  1e-2\\
NGC 6779 &  1.37 & 0.37 &  9.9 &  4.98 &  1.09 &  1.6\\
NGC 6229 &  1.61 & 0.13 & 29.3 &  5.35 &  1.14 &  2.6\\
NGC 6218 &  1.38 & 0.66 &  4.7 &  5.07 &  0.93 &  2.7\\
PAL 1    &  1.50 & 0.32 &  9.7 &  2.80 &  0.93 &  1e-2\\
NGC 6934 &  1.53 & 0.25 & 15.2 &  5.06 &  1.14 &  1.5\\
NGC 4833 &  1.25 & 1.00 &  5.9 &  5.20 &  1.77 &  1.2\\
NGC 6626 &  1.67 & 0.24 &  5.7 &  5.36 &  0.41 & 19.3\\
NGC 6637 &  1.39 & 0.34 &  8.2 &  5.30 &  0.84 &  5.6\\
NGC 6093 &  1.95 & 0.15 &  8.7 &  5.24 &  0.39 & 12.6\\
NGC 6528 &  2.29 & 0.09 &  7.4 &  4.67 &  0.20 & 10.4\\
NGC 6539 &  1.60 & 0.57 &  7.9 &  4.48 &  1.35 &  0.3\\
NGC 6517 &  1.82 & 0.06 & 10.5 &  4.88 &  0.19 & 26.3\\
NGC 6638 &  1.40 & 0.26 &  8.2 &  4.50 &  0.64 &  1.5\\
NGC 5824 &  2.45 & 0.05 & 31.3 &  6.00 &  0.47 & 36.5\\
NGC 6715 &  1.84 & 0.11 & 26.2 &  5.99 &  0.86 & 15.9\\
NGC 6171 &  1.51 & 0.54 &  6.3 &  4.80 &  1.02 &  1.0\\
\hline
\end{tabular}
\newpage
\begin{tabular}{|c|c|c|c|c|c|c|}
\hline
ID\hfil &$ c $\hfil&$r_c^{'}$\hfil &$R$\hfil &${\lg M/M_{\odot}}$\hfil &$R_c$\hfil&$\sigma_0$\\
\hline
\hline
NGC 6496 &  0.70 & 1.05 & 11.6 &  4.29 &  3.65 &  0.4 \\
NGC 6760 &  1.59 & 0.33 &  7.3 &  4.75 &  0.72 &  1.6 \\
NGC 5286 &  1.46 & 0.29 & 10.7 &  5.48 &  0.93 &  6.2 \\
NGC 6342 &  2.50 & 0.05 &  9.1 &  4.81 &  0.14 & 27.1 \\
NGC 6441 &  1.85 & 0.11 &  9.7 &  5.88 &  0.32 & 89.4 \\
NGC 6287 &  1.60 & 0.26 &  8.4 &  4.65 &  0.66 &  1.6 \\
NGC 6584 &  1.20 & 0.59 & 13.0 &  5.09 &  2.30 &  0.6 \\
NGC 1904 &  1.72 & 0.16 & 12.6 &  5.19 &  0.60 &  5.7 \\
TON 2    &  1.30 & 0.54 &  7.9 &  4.08 &  1.28 &  0.2 \\
NGC 1261 &  1.27 & 0.39 & 16.0 &  5.17 &  1.87 &  0.9 \\
NGC 6453 &  2.50 & 0.07 & 10.9 &  4.98 &  0.23 & 14.3 \\
NGC 5694 &  1.84 & 0.06 & 33.9 &  5.57 &  0.61 & 12.2 \\
NGC 6541 &  2.00 & 0.30 &  7.4 &  5.52 &  0.67 &  8.0 \\
NGC 6981 &  1.23 & 0.54 & 16.8 &  4.80 &  2.72 &  0.2 \\
NGC 6101 &  0.80 & 1.15 & 15.1 &  4.83 &  5.21 &  0.2 \\
NGC 6266 &  1.70 & 0.18 &  6.7 &  5.59 &  0.36 & 41.0 \\
HP 1     &  2.50 & 0.03 &  7.2 &  4.79 &  0.06 & 114.9\\
TER 5    &  1.74 & 0.18 &  8.0 &  4.94 &  0.43 &  6.2\\
NGC 5024 &  1.78 & 0.37 & 18.4 &  5.68 &  2.04 &  1.5 \\
NGC 6333 &  1.15 & 0.58 &  8.3 &  5.23 &  1.44 &  2.4 \\
NGC 104  &  2.04 & 0.37 &  4.3 &  6.03 &  0.48 & 49.0 \\
NGC 6355 &  2.50 & 0.05 &  7.1 &  4.79 &  0.11 & 42.6 \\
ERI      &  1.10 & 0.25 & 80.8 &  3.89 &  6.06 & 7e-3   \\
NGC 6553 &  1.17 & 0.55 &  4.7 &  5.42 &  0.78 & 12.2  \\
NGC 2808 &  1.77 & 0.26 &  9.3 &  5.96 &  0.73 & 22.4  \\
NGC 6256 &  2.50 & 0.02 &  9.3 &  4.52 &  0.06 & 83.3  \\
NGC 6366 &  0.92 & 1.83 &  3.6 &  4.47 &  1.98 &  0.4  \\
NGC 6388 &  1.70 & 0.12 & 11.5 &  6.16 &  0.41 & 116.4 \\
NGC 6652 &  1.80 & 0.07 &  9.4 &  4.98 &  0.20 & 30.9  \\
TER 6    &  2.50 & 0.05 &  7.5 &  4.53 &  0.11 & 20.9  \\
NGC 6642 &  1.99 & 0.10 &  7.6 &  4.16 &  0.23 &  3.0  \\
PAL 10   &  0.83 & 0.77 &  7.7 &  4.45 &  1.78 &  0.7  \\
NGC 6235 &  1.33 & 0.36 &  9.7 &  4.42 &  1.05 &  0.5  \\
NGC 6356 &  1.54 & 0.23 & 14.6 &  5.69 &  1.01 &  7.9  \\
NGC 6752 &  2.50 & 0.17 &  3.9 &  5.16 &  0.20 & 28.6  \\
TER 7    &  1.08 & 0.61 & 23.0 &  4.37 &  4.21 &  4e-2  \\
ARP 2    &  0.90 & 1.59 & 27.6 &  4.06 &  13.17&  4e-3  \\
NGC 6401 &  1.69 & 0.25 &  7.5 &  4.94 &  0.56 &  3.8  \\
PAL 11   &  0.69 & 2.00 & 12.6 &  4.82 &  7.56 &  0.4  \\
NGC 2298 &  1.28 & 0.34 & 10.6 &  4.56 &  1.08 &  0.7  \\
NGC 5466 &  1.43 & 1.96 & 16.6 &  4.85 &  9.76 &  1e-2  \\
NGC 6522 &  2.50 & 0.05 &  7.0 &  4.97 &  0.11 & 66.3  \\
NGC 6544 &  1.63 & 0.05 &  2.5 &  4.53 &  0.04 & 356.5 \\
NGC 7006 &  1.42 & 0.24 & 40.7 &  5.13 &  2.93 &  0.3 \\
TER 8    &  0.60 & 1.00 & 25.4 &  4.56 &  7.62 &  0.2  \\
\hline
\end{tabular}
\newpage
\begin{tabular}{|c|c|c|c|c|c|c|}
\hline
ID\hfil &$ c $\hfil&$r_c^{'}$\hfil &$R$\hfil &${\lg M/M_{\odot}}$\hfil &$R_c$\hfil&$\sigma_0$\\
\hline
\hline
NGC 6749 &   0.83 & 0.77 &  7.7 &  4.45 &  1.78 &  0.7\\
PAL 3    &  1.00 & 0.48 & 89.4 &  4.36 &  12.87 & 6e-3\\
NGC 5946 &  2.50 & 0.08 & 12.3 &  4.88 &  0.30 &  6.8\\
USK 3    &  2.10 & 0.15 &  7.5 &  5.11 &  0.34 & 11.3\\
PAL 4    &  0.78 & 0.55 & 99.6 &  4.21 &  16.43&  7e-3\\
NGC 6717 &  2.07 & 0.08 &  7.1 &  4.34 &  0.17 &  7.7  \\
NGC 7492 &  1.00 & 0.83 & 25.2 &  3.90 &  6.27 &  8e-3\\
NGC 6838 &  1.15 & 0.63 &  3.8 &  4.29 &  0.72 &  1.1\\
NGC 6293 &  2.50 & 0.05 &  8.8 &  5.01 &  0.13 & 45.9\\
PAL 5    &  0.74 & 2.90 & 22.6 &  3.92 &  19.66&  4e-3\\
NGC 362  &  1.94 & 0.17 &  8.3 &  5.45 &  0.42 & 17.7\\
E 3      &  0.75 & 1.87 &  4.2 &  3.55 &  2.36 &  0.1\\
NGC 5904 &  1.87 & 0.40 &  7.3 &  5.66 &  0.88 &  7.1\\
PAL   2  &  1.45 & 0.24 & 26.9 &  4.95 &  1.94 & 0.4\\
PAL 6    &  1.10 & 0.66 &  6.7 &  4.81 &  1.33 &  1.2\\
NGC 5272 &  1.85 & 0.50 & 10.0 &  5.81 &  1.50 &  3.5\\
NGC 5927 &  1.60 & 0.42 &  7.4 &  5.32 &  0.93 &  3.7\\
NGC 6284 &  2.50 & 0.07 & 14.3 &  4.88 &  0.30 &  6.6\\
NGC 4590 &  1.64 & 0.69 & 10.1 &  4.95 &  2.09 &  0.3\\
NGC 6273 &  1.53 & 0.43 &  8.5 &  6.03 &  1.10 & 14.7\\
TER 1    &  2.50 & 0.04 &  6.5 &  3.51 &  0.08 &  4.2\\
NGC 2419 &  1.40 & 0.35 & 82.3 &  6.01 &  8.64 &  0.3\\
NGC 6681 &  2.50 & 0.03 &  8.7 &  4.85 &  0.08 & 90.4\\
TER 2    &  2.50 & 0.03 &  9.5 &  3.88 &  0.09 &  8.1\\
NGC 5053 &  0.82 & 2.25 & 16.2 &  4.41 &  10.94&  2e-2\\
NGC 6362 &  1.10 & 1.32 &  7.5 &  4.91 &  2.97 &  0.3\\
LIL 1    &  2.30 & 0.06 & 10.5 &  5.27 &  0.19 & 45.8\\
NGC 7099 &  2.50 & 0.06 &  7.9 &  4.91 &  0.14 & 31.5\\
NGC 4372 &  1.30 & 1.75 &  4.9 &  5.11 &  2.57 &  0.4\\
NGC 6121 &  1.59 & 0.83 &  2.2 &  4.83 &  0.55 &  3.5\\
PAL 8    &  1.53 & 0.40 & 12.4 &  4.90 &  1.49 &  0.6\\
RUP 106  &  0.70 & 1.00 & 20.6 &  4.83 &  6.18 &  0.5\\
TER 3    &  0.70 & 1.18 & 26.4 &  4.47 &  9.35 &  0.1\\
NGC 6352 &  1.10 & 0.83 &  5.6 &  4.57 &  1.39 &  0.6\\
NGC 288  &  0.96 & 1.42 &  8.1 &  4.64 &  3.45 &  0.2\\
NGC 6341 &  1.81 & 0.23 &  8.1 &  5.34 &  0.56 &  8.8\\
NGC 6440 &  1.70 & 0.13 &  8.0 &  5.63 &  0.31 & 60.5\\
NGC 7078 &  2.50 & 0.07 & 10.2 &  5.85 &  0.21 & 120.8\\
NGC 7089 &  1.80 & 0.34 & 11.4 &  5.78 &  1.16 &  5.6 \\
NGC 6809 &  0.76 & 2.83 &  5.3 &  5.03 &  4.50 &  0.7 \\
IC 4499  &  1.11 & 0.96 & 18.4 &  5.12 &  5.30 &  0.1 \\
NGC 6254 &  1.40 & 0.86 &  4.3 &  5.06 &  1.11 &  1.8 \\
NGC 6397 &  2.50 & 0.05 &  2.2 &  4.63 &  0.03 & 306.8\\
IC 1276  &  1.29 & 1.08 &  9.3 &  5.09 &  3.01 &  0.3 \\
NGC 6144 &  1.55 & 0.94 & 10.1 &  4.76 &  2.85 &  0.1 \\
 \hline
\end{tabular}                                       
\newpage
 Column 1-- the cluster number; column 2-- the concentration parameter $c = \lg(r_t/r_c),$ where $r_t$ is the outer cluster radius and $r_c$ is the core radius; column 3-- the core radius $r_c$ in arcminutes; column 4-- the distance to the Sun $R$ in kpc; column 5-- the cluster mass in solar masses $m = \lg(M/M_\odot);$ column 6-- the core radius $R_c$ in pc; and column 7-- the central surface density $\sigma_0$ in g/cm~$^2.$

We took the data on $(ID), (c), (r_c^{'}),$ and $(R)$ from Harris (1996) and on ($\lg(M/M_\odot)$) from Mandushev et al. (1991). We calculated $(R_c)$ from $(r_c)$ and $(R)$ and $(\sigma_0)$ from formulas of the King model. The mass of an object in the King model is dened as
$$M=\pi r_c^2\sigma_0 \Lambda_0\eqno(1)$$
where
$$\Lambda_0=2\ln(r_t/r_c)-3\eqno(2)$$ 

For $r_t/r_c\sim 100,$ the surface density can be calculated with a high accuracy from the formula 
$$\sigma_0=\frac{10^{-5}M/M_{\odot}}{r_c^2\hbox{[pc]}}\eqno(3)$$
\v
\v
{\it The Surface-Density Function}
\v
The results of Table 1 reveal that $\sim 72\%$ of the Milky Way GCs have central surface densities up to 10 g/cm~${}^2.$ Figure 1 shows an empirical GC surfacedensity distribution (histogram 1). The sharp decrease at low $\sigma_0$ and the long tail at high $\sigma_0$ suggests that the distribution of this quantitymay be lognormal (histogram 2 in Fig. 1). The distribution of $\sigma_0$ is then $$f_{\d{\sigma_0}}(x)=\frac{1}{x\sigma\sqrt 2\pi}e^{-\d{\frac{{(\ln x-\mu)}^2}{2\sigma^2}}}\eqno(4)$$
where $\mu,\sigma$  are the distribution parameters and $x=\sigma_0.$ The values of $\mu=0.58,\hskip10pt \sigma=2.68$ were determined by the maximum--likelihood method.

To test the hypothesis of a lognormal distribution of the central surface density, we used the $\chi^2$ test. According to this test, the GC data are consistent with the hypothesis of a lognormal $\sigma_0$ distribution with $95\%$ confidence (at an $\alpha=5\%$ significance level.) Figure 2 shows the cumulative distribution function of the GC surface density. Curve 1 represents the lognormal distribution with the above parameters and curve 2 represents the experimental dependence for 135 objects.
For comparison, this figure also shows curve 3 for the exponential distribution function $\beta e^{-\beta x},$ where $ x=\sigma_0,$ with $\beta=0.056.$ Our analysis leads us to conclude that the lognormal distribution well describes the experimental data on the central surface density of Milky Way GCs. Our signicantly expanded [compared to the sample of Baryshev and Ezova (1997)] sample of GCs conrmed that the distribution function of the GC central surface density is lognormal and allowed the parameters of this distribution to be rened. How to physically interpret this law remains an open question, i.e., whether it is the result of selection eects or it is inherent in the physical formation and evolution of GCs.
\v
\v
\v

{\it The Critical Central Surface Density of a Lens}
\v
The central surface density of the putative lenses considered in the previous section is involved in many gravitational lensing formulas.

The main manifestations of gravitational focusing include the possible amplication of the image of a distance source. A signicant amplication factor is possible only if the observer is not between the lens focus and the lens. Since the quasar brightness is assumed to be signicantly amplied, by $3-5$ magnitudes, when interpreting quasar---galaxy associations, the position of the lens focus relative to the observer becomes important. In turn, the focus position is determined by the central surface density of the gravitational lens. Thus, we can determine the minimum lens density that corresponds to the case where the observer is at the lens focus (and, hence, a large image amplication is achieved).
$$\sigma_{0cr}(a,z_S)=\frac{0.077}{a(1-a)z_S}\eqno(5)$$ 
Assume that the quasars are located at the distance corresponding to the redshift $z_S = 2.$ A plot of this dependence for $z_S = 2$ is shown in Fig.3. We see from the gure that, if the lenses in quasar---galaxy pairs tend to be located at distances $a < 0.1$ and $a > 0.9$ [as follows from Bukhmastova (2001) and Baryshev and Ezova (1997)], then this corresponds to a lens surface density $\sigma_0>0.42$~ g/cm~$^2.$ An upper limit can be determined if we set $a = 10^{-4}$ and $z_S = 2.$ In that case, $\sigma_0 < 385$~g/cm~$^2,$ in close agreement with the central surface density for GCs (Table 1). Thus, the fact that the required central surface densities of putative gravitational lenses lie within the range of their values for GCs is one of the arguments that GCs are good candidates for gravitational lenses. The number of quasar--galaxy pairs found previously (Bukhmastova 2001) is 8382. These pairs include 3164 quasars. Assume that each quasar in associations is lensed and that GCs in the halos of galaxies in associations act as the lenses. Given the distances to the galaxy and the quasar, we can then calculate the critical central densities of the putative GC lenses for each pair that correspond to the condition that the observer be near the focus (and, hence, a large amplification factor is achieved). If our assumption is valid, then at least 3164 calculated critical densities must be characteristic of the GC central surface densities shown in Fig.1. Below, we present the calculated critical central surface densities of the putative lenses for 8382 quasar---galaxy pairs. The pairs can be found at http://www.astro.spbu.ru/staff/Baryshev/gl\_dm/htm.  The top line gives the critical density in g/cm~${^2}$ calculated from formula (5); the botton line gives the number of lenses with this density:
\v
\begin{tabular}{|c|c|c|c|c|c|c|c|c|c||c|}
\hline
 0-10& 10-20& 20-30& 30-40& 40-50& 50-60& 60-70& 70-80& 80-90& 90-100& >100\\
\hline
 708&  688&   556&   375&   272&   226&   295&   212&   470&   396&   4184\\
\hline
\end{tabular}
\v
 It turns out that 8332 pairs give central surface densities up to 80 g/cm~${^2},$ which closely match the central surface densities for GCs. This is yet another argument for the theory of the gravitational lensing of AGNs by GCs in galactic halos.

Dwarf galaxies, like GCs, also belong to the objects with a King mass distribution. They are difficult to detect because of their low luminosities. The data on dwarf galaxies (Ferguson and Binggeli 1994; Gallagher and Wyse 1994) are too scarce to reach a final conclusion about their central surface density. Dwarf galaxies contain a large amount of hidden mass, which leads to disagreements regarding the determination of the core radius. The parameters of dwarf galaxies from Yakovlev et al. (1983) give an idea of their basic parameters:

\v

\begin{tabular}{|c|c|c|c|c|}
\hline
\hfil& $\lg M/M_\odot$\hfil& $r_c$ \hfil& $\lg r_t/r_c$\hfil& $\sigma_0$\\
\hfil& \hfil& [kpc]\hfil &\hfil& [g cm$^2$]\\
\hline
dwarf galaxies & 5.0--9.5& 0.2--1& 0--0.7& $10^{-4}$--0.2\\
\hline
\end{tabular}
\v

If dwarf galaxies with characteristic surface densities $10^{-4}-0.2$~g/cm~$^2$ acted as the lenses, then an excess of pairs with $a\sim 0.5$ would be observed in the best case; i.e., galaxies would prefer the central position on the observer--quasar segment, which is in conict with the observational data from the catalog of quasar--galaxy associations. According to these data, $a < 0.1$ or $a > 0.9.$ Thus, for dwarf galaxies to be considered as good candidates for gravitational lenses, they must contain a large amount of hidden mass. The latter must be strongly concentrated toward the center in order to increase the central surface density to $1-80$~g/cm~$^2.$

If clustered transparent hidden mass objects act as the lenses, then we can specify the boundaries of their central surface densities: $\sigma_0 > 0.28 (z_Q\sim 3), \sigma_0 > 1.68 (z_Q\sim 0.5).$ The higher is the central surface density of a lens, the closer it is to the observer or the quasar. If the halo objects of our Galaxy (or the halo objects of the galaxy in which a quasar may be located) are assumed to be the lenses, then the central surface densities of these lenses must be $\sim 10^5-10^8$~g/cm~$^2,$ which correspond to the parameters of actually existing objects-- asteroids and minor planets. At present, such objects are dicult to invoke as the gravitational lenses to explain quasar—galaxy associations, because the model of a pointlike lens should most likely be used for them. For the lensing probability in this model to be high enough, the halo mass must be too large.
\v
\v
{\it The Caustics and Cross Sections of Gravitational Lenses}
\v
If we use the models of a pointlike gravitational lens applicable to stars or of an isothermal sphere applicable to galaxies to explain the enhanced brightness of the quasars from associations, then an accurate alignment of the observer, the lens, and the source on one straight line (on the axial caustic) is required to achieve large amplication factors.

An important feature of the gravitational lenses with a King mass distribution is that in this case, no accurate alignment of the observer, the lens, and the source on one straight line is required to achieve large amplication factors. This is because large amplication factors are achieved near the caustics. In addition to the axial caustic, GC lenses have the conical caustic. Its presence and the wide range of possible GC central surface densities ensure great freedom of the observer's position relative to the lens and the source and, hence, increase the probability of strong lensing [see Baryshev and Ezova (1997) for more details].

The probability that a distant object will be amplied by a foreground lens depends on the lens cross section. The larger is this cross section, the higher is the probability. For lenses with a King mass distribution, there is the total cross section that consists of the cross sections of the two (axial and conical)caustic regions. Therefore, the following two regions of maximum cross section exist for lenses with a King mass distribution: region I where all caustics meet (near--focus region) and region II on the line of sight halfway between the observer and the lens. This is schematically shown in Fig. 4. To take into account region II requires an accurate alignment of the observer, the lens, and the source on one straight line. In addition, as was mentioned above, a much lower central surface density than the GC surface density is required for the lenses located halfway between the observer and the source. To take into account region I, we need not impose stringent requirements on the observer's position relative to the source--lens straight line. Besides, the GC central surface densities exactly correspond to the case where the observer falls within region I, which explains the preferential positions of the galaxies in associations close to the observer and the quasar.

Thus, the presence of a conical caustic in GC lenses gives them yet another advantage over other possible objects.
\v
\v
AROUND WHAT TYPES OF GALAXIES

SHOULD GC LENSES BE SEARCHED FOR?
\v
The new catalog of quasar---galaxy associations contains 8382 pairs, which comprise 3164 quasars and 1054 galaxies. If galactic halo GCs are considered as the lenses located in the halos of these galaxies, then the following question naturally arises:
what types of galaxies are encountered more often in pairs? or what galaxies contain GCs around them and at what distances from the galactic center should GCs be searched for?

The original LEDA catalog of galaxies (Paturel 1997), which was used to compose pairs, contains data on the types of 40475 galaxies. Types -5, 5, and 10 correspond to elliptical, spiral, and irregular galaxies, respectively. Table 2 contains data on the number of galaxies of each type:
\v

\begin{tabular}{|c|c|c|c|c||c|}
\hline
 1&    2&          3&      4&    5&      6\\
\hline 
-5&    2500&       0.061&  37&   0.048&  0.79 \\
-4&    1474&       0.036&  15&   0.019&  0.54 \\
-3&    2872&       0.070&  34&   0.044&  0.63 \\  
-2&    3295&       0.081&  48&   0.063&  0.77 \\   
-1&    1554&       0.038&  24&   0.031&  0.82 \\
 0&    2053&       0.050&  30&   0.039&  0.78 \\
 1&    1968&       0.048&  25&   0.033&  0.67 \\
 2&    2327&       0.057&  32&   0.042&  0.73 \\
 3&    3967&       0.098&  58&   0.076&  0.78 \\
 4&    5232&       0.129&  57&   0.075&  0.58 \\
 5&    5995&       0.148&  51&   0.067&  0.45 \\
 6&    2730&       0.067&  55&   0.072&  1.07 \\
 7&    1053&       0.026&  38&   0.050&  1.92 \\
 8&    1171&       0.028&  50&   0.066&  2.88 \\
 9&     918&       0.022&  67&   0.088&  3.90 \\
10&    1366&       0.033&  136&  0.179&  5.32 \\
\hline
\end{tabular}
\v

column 1-- the type of galaxy; column 2-- the number of galaxies of a given type in the original LEDA catalog of galaxies;column 3-- the fraction of the galaxies of a given type from the total number of galaxies; column 4--the number of galaxies of a given type in quasar--galaxy pairs; column 5-- the fraction of the galaxies in pairs from the total number of galaxies in pairs for which data on their types are available (there are 757 such galaxies); column 6 reflects the relative contribution of the galaxies of each type in pairs. We see from the sixth column of this table that the relative number of GC lenses in the galaxies of types -5---5 is approximately the same. Starting from type 6, the relative number of putative lenses increases; in the irregular galaxies of type 10, the number of GCs is expected to be approximately a factor of 7 larger than that in the spiral and elliptical galaxies of each type.

Table 3 presents irregular galaxies of types 9 and 10 in quasar--galaxy associations:

column 1-- the galaxy name; column 2-- the galaxy type; column 3-- the galaxy redshift; column 4-- the assumed distances (in kpc) from the galactic center at which the clustered object (CO) lenses are located; column 5-- the number of quasars projected onto the galaxy; and column 6-- the number of compact star clusters detected in these galaxies to date
\v

\begin{tabular}{|c|c|c|c|c|c|}
\hline
1&            2&   3&        4&          5&  6\\
\hline
PGC0009168&  9 &   0.0019&   134&   1         &      \\
PGC0048280&  10&   9.0E-4&   42-149 &   49    &        \\
PGC0043869&  10&   0.0013&   58-149 &   27    &          \\
PGC0037722&  10&   0.0038&   17  &   1        &            \\
PGC0028623&  10&   0.0047&   133&   1         &              \\
PGC0039619&  9 &   7.0E-4&   11 &   48        &                \\
PGC0071596&  9 &   0.0014&   10  &   1        &                  \\
PGC0022151&  10&   0.0167&   133&   1        &\\
PGC0020852&  10&   0.0015&   127&   1         & \\
PGC0039615&  10&   9.0E-4&   23-147 &   9     &   \\
PGC0063000&  10&   5.0E-4&   84-137 &   2     &     \\
PGC0009759&  10&   0.0019&   94  &   1        &       \\
PGC0045755&  10&   0.0028&   88-149 &   2     &         \\
PGC0030187&  9&    0.0022&   104-133&   3     &           \\
PGC0042832&  10&   0.0015&   52  &   1        &             \\
PGC0003085&  9 &   6.0E-4&   63-142 &   7     & 2              \\
PGC0039018&  9 &   4.0E-4&   12-149 &   84    &                 \\
PGC0003710&  9 &   0.0046&   114&   1         &                   \\
PGC0010682&  10&   0.0036&   91  &   1         &\\ 
PGC0023769&  10 &  0.0013&   96-131 &   3   & \\
PGC0041608&  10&   6.0E-4&   6-147  &   88  &   \\
PGC0011153&  9 &   0.0059&   53&  1         &     \\
PGC0050073&  10&   0.0017&   142&   1  &            \\          
PGC0039225&  10&   0.0009&    102-144 & 4& 29   \\
PGC0016282&  pec(-3)&0.0021&  12-35&2&  16       \\
PGC0046039&  10& 0.0006& 40-140&  18  & 3         \\ 
\hline
\end{tabular}
\newpage
\begin{tabular}{|c|c|c|c|c|c|}
\hline
1&            2&   3&        4&          5&  6\\
\hline
PGC0006430&  10&   0.0013&   112-136&    3&     \\  
PGC0004143&  10&   0.0062&   75-148&      5&      \\
PGC0057170&  10&   0.0407&   46&       1    &       \\
PGC0044681&  10&   0.0028&   39&       1     &        \\
PGC0038148&  10&   7.0E-4&   8-141&   11     &          \\
PGC0014664&  10&   0.0029&   90&   1         &        \\
PGC0049497&  10&   0.0032&   98-125&  2      &         \\
PGC0042393&  9&    0.0041&   77-119 & 2      &          \\
PGC0005208&  10&   0.0072&   86-105 & 4       &          \\
PGC0054648&  10&   0.0022&   78  & 1         &            \\
PGC0020445&  10&   0.0011&   119& 1           &            \\
PGC0021073&  10&   0.0012&   50-129 & 2       &             \\
PGC0043450&  10&   0.0017&   76-144 & 4       &        \\
PGC0010670&  9 &   0.0037&   20-94  & 3       &         \\
PGC0012762&  9 &   0.0048&   46  & 1          &          \\
PGC0045939&  10&   5.0E-4&   24-142 & 20      &           \\
PGC0027935&  10&   0.0018&   85-85  & 1       &            \\
PGC0002992&  10&   0.0022&   131& 1           &             \\
PGC0014458&  9 &   0.0063&  41  & 1           &             \\
PGC0013985&  10&   0.0029&   120-148& 2       &   \\
PGC0001014&  9&   4.0E-4 &  7-149   & 100     &    \\
PGC0029034&  9 &   0.0052&   93.6  & 1        &     \\
PGC0017373&  10&   0.0062&   106& 1           &      \\
PGC0051281&  10&   0.0024&   105-137& 2       &       \\
PGC0029194&  10&   0.0012&   75-145 & 2       &        \\
PGC0031427&  10&   0.0099&  39  & 1           &        \\
PGC0016344&  9 &   0.0056&   134& 1           &          \\
PGC0011139&  9 &   0.0019&   84  & 1          &           \\
PGC0038881&  10&   5.0E-4&   71-144 & 11      &            \\
PGC0029427&  10&   0.0021&   61-148 & 6       &             \\
PGC0033479&  10&   0.0113&   146& 1           &   \\
PGC0036398&  9 &   0.0044&   61  & 1          &    \\
PGC0006574&  9 &   0.0011&   95  & 1          &     \\
PGC0042219&  9 &   0.0025&   45-142 & 10      &      \\
PGC0052142&  10&   7.0E-4&  86-141 & 2        &      \\
PGC0041746&  9 &   9.0E-4&   10-144 & 39      &        \\
PGC0026663&  10&   0.0061&   40-44  & 2        &        \\
PGC0045927&  10&   0.0032&   62  & 1          &          \\
PGC0041743&  9 &   0.0035&   116-129& 2       &           \\
PGC0024213&  10&   5.0E-4&   53-137 & 2        &           \\
PGC0004913&  9 &   0.0021&   110& 1             &           \\
PGC0051472&  10&   5.0E-4&   51-149 & 29        &            \\
PGC0038481&  10&   0.0029&   50-141 & 3         &             \\
PGC0067908&  10&   4.0E-4&   89-121 & 6         &              \\
PGC0027091&  10&   0.0019&   10  & 1            &     \\
\hline
\end{tabular}
\newpage
\begin{tabular}{|c|c|c|c|c|c|}
\hline
1&            2&   3&        4&          5&  6\\
\hline
PGC0010854&  9 &   0.0050&   130-144& 2         &      \\
PGC0023340&  9 &   0.0017&   99& 1              &       \\
PGC0009685&  10&   0.0047&   137& 1             &        \\
PGC0007399&  10&   0.0023&   139& 1             &         \\
PGC0003844&  10&   8.0E-4&   6-149  & 47        &          \\
PGC0031414&  10&   0.0032&   144&  1            &           \\
PGC0008851&  9 &   0.0054&   110& 1             &            \\
PGC0068498&  10&   0.0061&   112& 1             &             \\
PGC0072228&  9 &   9.0E-4&   57-129 & 5         &              \\
PGC0010650&  10&   0.0089&   109-120& 4         &              \\
PGC0032041&  9 &   0.0027&   18  & 1            &               \\
PGC0002816&  10&   0.0057&   59-65  & 2         &                \\
PGC0005896&  9 &   0.0013&   20-149 & 79        &    \\
PGC0001357&  10&   0.0118&   55  & 1            &     \\
PGC0039142&  10&   9.0E-4&   17-149 & 37        &      \\
PGC0037050&  10&   8.0E-4&   33-130 &  9        &       \\
PGC0043822&  10&   0.0142&   83  &  1           &        \\
PGC0039734&  9 &   0.0016&   19  &  1           &         \\
PGC0033264&  10&   0.0023&   17-120 &  3        &          \\
PGC0002578&  10&   0.0012&   26-138 &  3        &           \\
PGC0017157&  9 &   0.0043&   126& 1            &             \\
PGC0042160&  10&   0.0036&   43-139 &  2        &             \\
PGC0006751&  10&   0.0057&   81  &  1           &              \\
PGC0045314&  9 &   0.0011&   44-149 &  10       &               \\
PGC0047368&  9 &   7.0E-4&   73-149 &  11       &                \\
PGC0049452&  10&  8.0E-4 &  38-137  &  2        &    \\
PGC0009702&  10&   0.0019&   147&  1            &     \\
PGC0023324&  10&   5.0E-4&   84-149 &  3        & 3     \\
PGC0021199&  10&  0.0018 &  142 &  1            &       \\
PGC0032226&  9&    0.0019&   86  &  1           &        \\
PGC0045506&  10&   7.0E-4&   31-144 &  14        &        \\
PGC0047951&  9 &  0.0048 &  95 &   1            &          \\
PGC0032221&  10&   0.0026&   40  &   1          &           \\
PGC0037037&  10&   0.0030&   24-140 &    3      &            \\
PGC0045306&  10&   0.0039&   12-147 &   8        &            \\
PGC0059410&  10&   0.0015&   96  &  1            &             \\
PGC0000703&  10&   0.0030&   127&   1            &              \\
PGC0071538&  10&   6.0E-4&   38-148 &   2        &               \\
PGC0042348&  10&   0.0069&   110&   1            &                \\
PGC0049448&  10&   5.0E-4&   17-142 &   25       &    \\
PGC0021585&  10&   0.0012&   45-123 &   3        &     \\
PGC0041512&  10&   0.0021&   108-144&   6        &      \\
PGC0057678&  9 &   0.0024&   61-126 &   2        &       \\
PGC0040844&  10&   0.0011&   25-148 &  25        &        \\
PGC0014377&  10&   0.0063&   92  &   1           &         \\
\hline
\end{tabular}
\newpage
\begin{tabular}{|c|c|c|c|c|c|}
\hline
1&            2&   3&        4&          5&  6\\
\hline
PGC0015996&  10&   0.0045&   122&   1            &          \\
PGC0039317&  10&   0.0039&   47-136 &   6        &           \\
PGC0032376&  9 &   0.0045&   44-111 &   4        &            \\
PGC0067048&  10&   0.0083&   35-143 &   4        &             \\
PGC0039316&  10&   6.0E-4&   18-137 &   16       &              \\
PGC0070306&  9 &   0.0046&   50  &   1           &               \\
PGC0046127&  10&   9.0E-4&   11-146 &   18       &    \\
PGC0003974&  9&   5.0E-4 &  73-141  &   3        &     \\
PGC0029549&  10&   0.0046&   56-127 &   2        &      \\
PGC0044982&  10&   0.0025&   118-132&   2        &       \\
PGC0050961&  10&   5.0E-4&   56-143 &   14       &        \\
PGC0035684&  10&   8.0E-4&   54-131 &    2       &         \\
PGC0041505&  9 &   9.0E-4&   18-145 &   35       &          \\
PGC0041667&  10&   0.0109&   17  &   1           &           \\
PGC0014407&  10&   0.0047&   58  &   1           &            \\
PGC0005661&  10&   0.0022&   17  &   1           &             \\
PGC0025754&  9 &   0.0164&   122&   1            &              \\
PGC0037217&  9 &   0.0033&   109&   1           &                \\
PGC0038998&  10&   5.0E-4&   58-145 &   16      &      \\
PGC0012706&  10&   0.0046&   97  &   1          &       \\
PGC0035878&  10&   8.0E-4&   36-142 &   8       &        \\
PGC0040791&  10&   7.0E-4&   21-148 &   6       &         \\
PGC0031693&  10&   0.0024&   146&   1           &          \\
PGC0071878&  9 &   0.0092&   34-115 &   2       &           \\
PGC0040596&  9 &   0.0011&   1-112  &    4      &            \\
PGC0037967&  9 &   0.0055&   127-141&   2       &             \\
PGC0055965&  10&   0.0070&   93  &   1          &              \\
PGC0021600&  10&   9.0E-4&   37  &     1        &               \\
PGC0039260&  9 &   0.0029&   70-134 &      5    &      \\
PGC0010213&  9 &   0.0046&   100&   1           &       \\
PGC0046666&  10&   0.0082&   128-134&   2       &       \\
PGC0042910&  9 &   0.0039&   131&   1           &        \\
PGC0041258&  10&   0.0016&   54-116 &    10     &         \\
PGC0038190&  10&   0.0049&   114&   1           &          \\
PGC0013521&  9 &   0.0050&   102-147&   3       &           \\
PGC0044170&  10&   0.0019&   93  &   1          &            \\
PGC0069271&  9 &   0.0047&   109&   1           &             \\
PGC0029128&  9 &   0.0013&   98-141 &   3       &              \\
PGC0069505&  10&   0.0059&   12-130 &   5       &               \\
PGC0071466&  9 &   0.0023&   114 &  1           & \\
PGC0042275&  10&   0.0010&   54-128 &   4       &  \\
PGC0037553&  10&   0.0036&   24-107 &   2       &   \\
PGC0040973&  10&   7.0E-4&   68-145 &   7       &    \\
PGC0042901&  10&   0.0020&   64-149 &   9       &     \\
PGC0033010&  10&   0.0023&   136&   1           &      \\
\hline
\end{tabular}
\newpage
\begin{tabular}{|c|c|c|c|c|c|}
\hline
1&            2&   3&        4&          5&  6\\
\hline
PGC0032343&  10&   0.0025&   118-146&   2       &       \\
PGC0038811&  9 &   0.0033&   44-106 &   3       &        \\
PGC0004416&  10&   0.0039&   77  &   1          &         \\
PGC0072525&  10&   0.0022&   98-134 &   2       &          \\
PGC0042863&  9 &   0.0021&   101-133 &  2       &           \\
PGC0000035&  10&   0.0011&   27-64  &   2       &            \\
PGC0035856&  10&   8.0E-4&   52-149 &   4       &             \\
PGC0033408&  9 &   0.0023&   71-138 &   3       &    \\
PGC0031477&  10&   0.0020&   142&   1           &     \\
PGC0042503&  9 &   0.0021&   43-120 &   6       &      \\
PGC0045388&  9 &   0.0159&   10-133 &   2       &       \\
PGC0004211&  10&   0.0049&   68-138 &   4       &        \\
PGC0017652&  10&   0.0037&   102&   1           &         \\
PGC0041636&  10&   0.0011&   53-124 &    5      &          \\
PGC0000621&  10&   7.0E-4&   38-140 &   6       &      \\
PGC0035451&  9 &   0.0049&   113-127&   2       &       \\
PGC0037542&  9 &   0.0032&   83  &   1          &        \\
PGC0033957&  9 &   0.0050&   141&   1           &         \\
PGC0041239&  10&   0.0013&   48-133 &   14      &          \\
PGC0040409&  10&   0.0029&   75  &   1          &           \\
PGC0039431&  10&   5.0E-4&   7-148  &   96      &            \\
PGC0030997&  9 &   6.0E-4&   91-143 &   4       &             \\
PGC0043129&  10&   0.0011&   106-142&   5       &               \\
PGC0048332&  10&   7.0E-4&   40-131 &   6       &      \\
PGC0042656&  10&   0.0012&   69-121 &   3       &       \\
PGC0041036&  10&   0.0019&   50-132 &   4       &        \\
PGC0040563&  10&   0.0050&   47  &   1          &         \\
PGC0039395&  10&   0.0043&   53-132 &   3       &           \\
PGC0043124&  10&   0.0017&   128-142&    2      &             \\
PGC0040367&  9 &   0.0016&   97  &   1          &               \\
PGC0004202&  10&   0.0021&   91  &   1        &                   \\
PGC0040200&  9 &   0.0013&   110-135&   2     &                     \\
PGC0040005&  10&   5.0E-4&   23-148 &   103   &                       \\
PGC0056959&  10&   0.0027&   87  &   1        &                         \\
PGC0027605&  10&   5.0E-4&   48-141 &   18    &    \\
PGC0049158&  10&   6.0E-4&   12-145 &   17    &     \\
PGC0026142&  10&   0.0020&   84  &   1        &      \\
PGC0003290&  9 &   0.0052&   93-147 &   8     &        \\
PGC0003095&  9 &   0.0054&   87-128 &   2     &          \\
PGC0011202&  9 &   0.0051&   103&   1         &            \\
PGC0029653&  10&   0.0011&   32-106 &   3     &  1            \\
PGC0026895&  10&   0.0121&   105&   1         &                \\
PGC0037727&  10&   0.0037&   61-136 &   3     & \\
PGC0041020&  9 &   9.0E-4&   11-128 &   5     &  \\
PGC0044536&  9 &   0.0028&  60-140 &   7      &   \\
\hline
\end{tabular}
\v
The data of this tablemake it possible to check specific galaxies for the presence of clustered objects in their halos and, in particular, in the immediate vicinity of the quasars.

The absolute magnitudes of the galaxies of types 9 and 10 in pairs lie within the range from -11 to -21, with $\sim 80\%$ of them having absolute magnitudes from -15 to -19. The distances from the centers of the irregular galaxies at which one might expect GCs are $\sim 50-100$~kpc. To all appearances, these are young globular clusters.
The currently available observational data confirm our conclusion that the halos of irregular galaxies contain clustered objects with the masses typical of the GCs. Based on the Hubble Space Telescope (HST) data, Hunter et al. (2001) investigated the irregular galaxy NGC 1569. In this galaxy, they found 47 clusters whose core radii lie within the range typical of the GCs. The absolute magnitude of the galaxy is $M_V =-18.$ The parameters of two clusters in this galaxy are known: the core radii are 2.3 and 3.1 pc and the masses are $\sim 3\cdot10^5 M_\odot.$ These authors provided data on seven more irregular galaxies in which compact star clusters were detected. Sixty compact star clusters were identified in the irregular galaxy NGC 4449, 49 of which are presumed to be GCs. NGC4214 (PGC 039225) is an irregular galaxy with active star formation. Twenty nine compact star clusters were identified around this galaxy; Hunter et al. (2001) believe these clusters to be young. Similarly, three clusters were found around the galaxy DDO 50 (PGC 023324), 16 clusters were found around NGC 1705 (PGC 016282), three clusters were found around DDO 168 (PGC 046039), four clusters were found around DDO165 (PGC 045372), and one compact cluster was found in DDO 75 (PGC 029653) and in WLM. All these compact objects have radii $r < 15$~pc and $M_V$ brighter than $-10^m.$ Most of the star clusters are young with sizes and masses comparable to those of GCs. Thus, the number of GCs in irregular galaxies can be much larger than is generally thought to be. The irregular galaxies NGC 4214, DDO 50,
NGC 1705, DDO 168, and DDO 75 with compact star clusters detected in their halos investigated by Hunter et al. (2001) are members of the quasar---galaxy associations found previously (Bukhmastova 2001); several quasars are projected onto each of them.

Table 4 lists the spiral and elliptical galaxies in whose halos GCs and quasars have been detected to date: column 1--- the galaxy name; column 2---the galaxy type; column 3--- the number of quasars projected onto the galaxy; and column 4--- the number of GCs detected in the halo. The data on the presence of GCs were taken from astro-ph/0206140 for the first galaxy and from astro-ph/0112209 for the second, third, and fourth galaxies; for the remaining galaxies, we took the data from Harris (1996).
Thus, the data of Tables 3 and 4 support the hypothesis that the associations of distant quasars with nearby galaxies can be provided by compact star clusters (GCs), which act as gravitational lenses. Therefore, a more thorough analysis of the properties of these associations may give an independent scenario for the formation and evolution of globular clusters or may shed light on the nature of the clustered hidden mass in galactic halos.
\v
\v

\begin{tabular}{|c|c|c|c|}
\hline
1&            2&   3&        4    \\
\hline
PGC041297 (NGC4478)&  E2 &    3         & presence\\
PGC044324 (NGC4840)&  SABO&   1    &  $10^3-10^4$      \\
PGC043008 (NGC4673)&  E1-2&   1    &  $10^3-10^4$        \\
PGC044553&  E&      1    &  $10^3-10^4$          \\
PGC013344 (NGC1387)&  SO &2 & $ 385\pm 80 $              \\
PGC013418 (NGC1399) &  E1pec  &1& $ 5340\pm 1780 $               \\
PGC013433 (NGC1404) &  E1     &1& $ 950\pm 140 $                \\
PGC024930 (NGC2683) &  Sb&    4 & $ 310\pm 100 $        \\
PGC032226 (NGC3377) &  E5&    1 & $ 210\pm 50 $  \\
PGC032256 (NGC3379) &  E1&    6 & $ 260\pm 140 $  \\
PGC036487 (NGC3842) &  E3&    2 & $ 14000\pm 2500 $ \\
PGC039764 (NGC4278) &  E1&    2 & $ 1050\pm 120 $   \\
PGC039246 (NGC4216) &  Sb&    129& $ 620\pm 310 $    \\
PGC041327 (NGC4486) &  EO&    1 & $ 13000\pm 500 $        \\
PGC041968 (NGC4552) &  EO&    34 & $ 2400 $           \\
PGC042051 (NGC4564) &  E6&   3& $ 1000\pm 300 $           \\
PGC042628 (NGC4621) &  E5&   7 & $ 1900\pm 400 $         \\
PGC000218 (NGC7814) &  Sab&   1& $ 500\pm 160 $                 \\
\hline
\end{tabular}
\v
\v
\v

MAIN CONCLUSIONS AND OBSERVATIONAL TESTS
\v
As a result of our analysis of the properties of quasar---galaxy associations and according to the  assumptions about their possible nature made in the introduction, we favor GCs as good candidates for gravitational lenses located in the halos of galaxies in associations. Apart from the aforesaid, the following arguments support this assumption:

(1) GCs are actually existing observed objects. Their number in the halos of some galaxies can reach several thousand (Harris 1996). This provides a good possibility to use them as good candidates for gravitational lenses, because the more potential candidates for gravitational lenses are in the halo, the  larger is the total cross section of the lenses and, hence, the higher is the probability of lensing distant sources.

(2) The possible manifestations of gravitational lensing include source image splitting. For GCs described by the King model, one, two, or three images of a distant source can appear. The characteristic linear separation between the possible images is defined via the Chwolson--- Einstein ring radius 
$\xi=\d{\sqrt{2r_g\frac {R_{OL}R_{LS}}{R_{OS}}}}, $ 
where $R_{OL}, R_{LS}$ and $R_{OS}$  are the distances from the observer to the lens, from the lens to the source, and from the observer to the source, respectively; and $r_g$ is the gravitational radius of the lens. If the GC mass is assumed to be $10^6 M_{\odot},$ then for $z_{OL} = 10^{-3}-0.1$  (the redshifts of the galaxies in associations), $\xi$ varies between $10^{-2}$  and $0.4$~pc. The quasar sizes are very small ($\sim 10^{-3}$~ pc). Since the possible amplification factor for a source is equal to the ratio of the area of its image to the source area, GCs make an amplification by $3^m-10^m$ possible. This favorably distinguishes them from stars, which are often invoked as the lenses to explain quasar---galaxy associations. Stars described by the model of a pointlike lens (microlensing) give only a weak amplification of the source and cannot explain the enhanced quasar luminosity in this way.

(3) The GC central surface densities closely correspond to the condition that the observer be near the lens focus. This gives large amplification factors $3^m-5^m$ and, hence, allows us to explain the enhanced quasar luminosity and their leftward displacement in the Hubble diagram $z=f(m)$ for galaxies and quasars (see Bukhmastova 2001).

(4) The preferential GC distances from the galactic centers are less than 50~kpc. This is in agreement with the fact that most of the nearby quasars from associations with $z_G/z_Q>0.9$ are also projected  onto galactic halos with radii up to 50~kpc (Bukhmastova 2001).

(5) Since the critical central surface densities of the putative lenses calculated for quasar---galaxy associations match the GC central surface densities, we can assume that the association between distant quasars and nearby galaxies is provided by halo GCs.

(6) The expected light variability because of the motion of a GC as a whole has a time scale of more than a thousand years, because the time scale estimate is given by the expression 
$$t_{var}\sim\frac{R_{OL}\Theta_{Ch-E}}{v}\sim 10^3 \Bigl(\frac{M}{10^6M_\odot}\Bigr)^{1/2}\Bigl(\frac{D^{\prime}}{10\hbox{Mpc}}\Bigr)^{1/2}\Bigl(\frac{v}{10^3\hbox{km s}^{-1}}\Bigr)^{-1}\hbox{yr}$$
where  $D^{\prime}=\d{\frac{R_{LS}R_{OL}}{R_{OS}}}$-- is the reduced distance to the lenses and $v$ is the lens velocity. However, light variability on time scales of less than one year is possible because of microlensing by individual GC stars.

(7) Since GCs belong to the galactic halos, absorption lines can emerge in the spectra of quasars with $z_{abs}$ corresponding to this galaxy. In this case, an excess of systems with $z_{abs}/z_{em}>0.9$ or $z_{abs}/z_{em}<0.1$ must be observed. This theoretical assumption follows from the dependence (Baryshev and Ezova 1997) of the differential lensing probability on  $a=z_L/z_S,$ where $z_L$ and $z_S$ are the redshifts of the lens and the source, respectively. Note that if the distribution of galaxies on the line of sight is fractal, then this can also cause the number of systems with $a>0.9$ and $a<0.1$  to increase. Thus, gravitational lensing by objects with a King mass distribution and the fractal distribution of galaxies with their surrounding lenses can mutually enhance this effect. More specifically, in this case, galaxies in pairs with quasars will avoid the central position on the observer--- quasar line of sight, as confirmed by analysis of the catalogs of associations (Burbidge et al. 1990; Bukhmastova 2001).

(8) Based on the observed widths of the 21--cm absorption lines in the spectra of some quasars and assuming the clouds responsible for their appearance to be gravitationally bound, Komberg (1986) estimated the optical luminosities and sizes of these clouds. The luminosities, sizes, and masses were found to be typical of the GCs. These author also estimated the HI mass to be only a few solar masses. Even in old GCs one might expect such an amount of neutral hydrogen. It thus follows that the 21--cm lines can be detected in nearby GCs. The coincidence of the  21--cm absorption lines and the systems with absorption lines of metal ions may also stem from the fact that GCs, which can be gravitational lenses (thus, can be responsible for the enhanced luminosities of some quasars), fall on the line of sight.

(9) An increased number of GCs, a factor of 6--7 larger than that for spiral and elliptical galaxies of each type (Table 2), must presumably be observed in irregular galaxies of types 9 and 10. These GCs are presumed to be located at distances of $\sim 100$~kpc from the galactic centers and to have central surface densities $\sigma_0\sim 1\div 200$~g~cm~$^{-2}.$ This follows from our analysis of the types of galaxies in pairs and from our calculations of the critical surface density of the putative lenses for each quasar---galaxy pair (formula 5) as well as from the available HST observational data (Hunter et al. 2001).

(10) The Chandra observations revealed X-ray sources in six Galactic GCs. Thus, GCs can be detected in galactic halos by X-ray sources (Revnivtsev et al. 2002).

(11) Quasars in associations with $a>0.9,$ i.e., close to galaxies in redshift, are observable in the radio frequency range. This allows the possible splitting of the core structure in these quasars to be detected at a millisecond level. These quasars can be found at http://www.astro.spbu.ru/sta./baryshev/gl\_dm.htm.

It follows from our analysis of the data on 135 Milky Way GCs given in Table 1 that the GC distribution in central surface density is lognormal. I hope that all of the arguments listed above will help to explain the appearance of quasar---galaxy associations and to re.ne the nature of the candidates for gravitational lenses. Apart from GCs, the latter can include dwarf galaxies and clustered hidden mass objects.
\v
\v

ACKNOWLEDGMENTS
\v
I am grateful to Yu.V. Baryshev for a helpful discussion and D.S. Bukhmastov for technical help in preparing the publication.
\v
Translated by V.Astakhov
\v
\v
\v

REFERENCES
\v

\begin{enumerate}

\item Arp// Astrophys.J., 1966, v.148, p.321.
\item Arp// Quasars, Redshifts and Controversies, Interstelar Media, Berkelley, Ca., 1987.
\item Arp// Astron. and Astrophys., 1990, v.229, p.93.
\item J.M. Barnothy// Astron.J., 1965, v.70, p.666. 
\item J.M. Barnothy// Bul.A.A.S., 1974, v.6, p.212. 
\item M. Bartusiak// Astronomy, 1998, June, p.42.
\item Yu. V. Baryshev, Yu. L. Ezova, (Yu. L. Bukhmastova,) //Astron. Zh.,1997,v.74,p. 497 (Astron. Rep., 1997, v.41, p.  436); astro-ph/020648.
\item P.V. Bliokh and A.A. Minakov// Gravitational Lenses, (Naukova Dumka, Kiev, 1989).
\item Yu.L. Bukhmastova// Astron. Zh., 2001, v.78, p.675 (Astron. Rep.,2001, v.45); astro-ph/0206456.
\item G. Burbidge, A.Hewitt, J.V.Narlikar, P.Das Gupta// Astrophys. J.Suppl.Ser., 1990, v.74, p.675.
\item C.R. Canizares// Nature, 1981, v.291, p.620.
\item B. Carr// ARA A, 1994, v.32, p.531.
\item Y. Chu, X. Zhu, G. Burbidge, A. Hewitt // Astr.Ap., 1984, v.138, p.408.
\item N. Dalal, C.S. Kochanek// (2002); (astro-ph/0111456).
\item F.De. Paolis, G. Ingrosso, Ph. Jetzer, M. Roncadelli// Astrophys.J., 1998, v.500, p.59. 
\item H.C. Ferguson, B. Binggeli// Astron. and Astrophys. Rev., 1994, v.6, p.67.  
\item J.S. Gallagher, R.F.G. Wyse// Publication of the Astron. Society of the Pacific., 1994, v.106, p.1225. 
\item W.E. Harris// A.J., 1996, v.112, p.1487 ;

(http://www.physics. mcmaster.ca/Globular.html)
\item A. Hewitt, G. Burbidge// Ap.J.Suppl., 1980, v.43, p.57.
\item L. C. Ho and J. Kormendy//(2000); astro-ph/000367.
\item D.A. Hunter, O.H. Billet and B.G. Elmegreen,// (2001); (astro-ph/0112260).
\item A. Klypin et al.// Astrophys. J., 1999, v.516, p.516.
\item B. V. Komberg// Astrofizika , 1986, v.24, p.321.
\item J. Kormendy// (2000); (astro-ph/ 0007400).
\item J. Kormendy// (2000); (astro-ph/ 0007401).
\item J. Kormendy, L.C. Ho//(2000); (astro-ph/ 0003268).
\item V. G. Kurt and O. S. Ugol’nikov// Kosm. Issled., 2000, v.38, p.227.
\item G. Mandushev, N. Spassova, A. Staneva// Astron. and Astrophys., 1991, v.252, p.94.
\item B. Moore et al.// Astrophys. J., 1999, v.524, p.L19.
\item B. Moore et al.// (2001); (astro-ph/0106271).
\item J.-L. Nieto// Nature, 1977, v.270, p.411.
\item J.-L. Nieto// Astr.Ap., 1978, v.70, p.219.
\item J.-L. Nieto// Astron. Astrophys., 1979, v.74, p.152.
\item J.-L. Nieto, M. Seldner//Astr. Ap., 1982, v.112, p.321.
\item G. Paturel// Astrophys. J. Suppl. Ser., 1997, v.124, p.109.  
\item P.J.E.Peebles, B. Ratra// The Cosmological Constant and Dark Energy. (2002); (astro-ph/0207347).
\item M.G. Revnivtsev, S. P. Trudolyubov, and K. N. Borozdin//  Pis’ma Astron. Zh., 2002, v.28, p. 276. (Astron. Lett. 2002, v.28, p.237)
\item M. Rozyczka// Acta Astr., 1972, v.22, p.93.
\item V. Sahni and A. A. Starobinsky//  Int. J. Mod. Phys. D, 2000, v.9, p.373 ;astro-ph/9904398.
\item P. Schneider, J. Ehlers,E.E. Falco// ''Gravitational Lenses'',
Berlin-Heidelberg- New York, Spriger-Verlag (1992).
\item J. Stuart, B. Wyithe and A. Loeb// (2002); (astro-ph/0203116).
\item J.A. Tyson// Astron.J., 1986, v.92, p.691.
\item O.S. Ugol’nikov// Preprint of Lebedev Physical Institute, Russian Academy of Sciences, Moscow , 2001, no. 17.
\item M.P. Veron-Cetty, P. Veron// Quasars and Active Galactic Nuclei, ESO Scientific Report 18 (1998).
\item D.G. Yakovlev, L.G. Mitrofanov, S.A. Levshakov and D.A. Varshalovich// Astrophis. and Space Science., 1983, v.91, p.133. 
\item A.V. Yushcenko, Yu. Baryshev, A.A. Raikov// Astron. Astrophys. Trans., 1998, v.17, p.9.
\item A. F. Zaharov// Gravitational Lenses and Microlenses (Yanus-K, Moscow, 1997).
\end{enumerate}

\end{document}